\documentclass[titlepage,11pt,twoside]{article}\usepackage[]{graphicx}\usepackage[]{color}
\makeatletter
\def\maxwidth{ %
  \ifdim\Gin@nat@width>\linewidth
    \linewidth
  \else
    \Gin@nat@width
  \fi
}
\makeatother

\definecolor{fgcolor}{rgb}{0.345, 0.345, 0.345}

\usepackage{framed}
\makeatletter
 {\par\unskip\endMakeFramed%
 \at@end@of@kframe}
\makeatother

\definecolor{shadecolor}{rgb}{.97, .97, .97}
\definecolor{messagecolor}{rgb}{0, 0, 0}
\definecolor{warningcolor}{rgb}{1, 0, 1}
\definecolor{errorcolor}{rgb}{1, 0, 0}
\newenvironment{knitrout}{}{} 

\usepackage{alltt}
\usepackage{natbib}
\usepackage{mathtools}
\usepackage[myheadings]{fullpage}
\usepackage{pmetrika}
\usepackage{submit}
\usepackage{amsmath,amssymb}
\usepackage[shortlabels]{enumitem}

\newcommand{\E}{\mathbb{E}}

\newcommand{\R}{\mathbb{R}}

\renewcommand{\P}{\mathbb{P}}

\IfFileExists{upquote.sty}{\usepackage{upquote}}{}
\begin{document}

\begin{titlepage}
\linespacing{1}

\title{A Robust Effect Size Index}

\author{Simon Vandekar}
\author{Ran Tao}
\author{Jeffrey Blume}

\affil{Department of Biostatistics, Vanderbilt University}
~\\
Please address correspondence to: \\
Simon Vandekar\\
2525 West End Ave., \#1136\\
Department of Biostatistics\\
Vanderbilt University\\
Nashville, TN 37203\\
{\tt simon.vandekar@vanderbilt.edu}

\linespacing{1}


\begin{center}\vskip3pt

\newpage

Abstract\vskip3pt

\end{center}
\begin{abstract}
Effect size indices are useful tools in study design and reporting because they are unitless measures of association strength that do not depend on sample size.
Existing effect size indices are developed for particular parametric models or population parameters.
Here, we propose a robust effect size index based on M-estimators.
This approach yields an index that is very generalizable because it is unitless across a wide range of models.
We demonstrate that the new index is a function of Cohen's $d$, $R^2$, and standardized log odds ratio when each of the parametric models is correctly specified.
We show that existing effect size estimators are biased when the parametric models are incorrect (e.g. under unknown heteroskedasticity).
We provide simple formulas to compute power and sample size and use simulations to assess the bias and variance of the effect size estimator in finite samples.
Because the new index is invariant across models, it has the potential to make communication and comprehension of effect size uniform across the behavioral sciences.

\end{abstract}

\vspace{\fill}
\end{titlepage}\vspace*{24pt}



\section{Introduction}
Effect sizes are unitless indices quantifying the association strength between dependent and independent variables.
These indices are critical in study design when estimates of power are desired, but the exact scale of new measurement is unknown \citep{cohen_statistical_1988}, and in meta-analysis, where results are compiled across studies with measurements taken on different scales or outcomes modeled differently \citep{chinn_simple_2000,morris_combining_2002}.
With increasing skepticism of significance testing approaches \citep{trafimow_null_2017,wasserstein_asas_2016,harshman_there_2016,wasserstein_moving_2019}, effect size indices are valuable in study reporting \citep{fritz_effect_2012} because they are minimally affected by sample size.

Effect sizes are also important in large open source datasets because inference procedures are not designed to estimate error rates of a single dataset that is used to address many different questions across tens to hundreds of studies.
While effect sizes have similar bias to $p$-values when multiple hypotheses are considered, obtaining effect size estimates for parameters specified {\it a priori} may be more useful to guide future studies than hypothesis testing because, in large datasets, $p$-values can be small for clinically meaningless effect sizes.

There is extensive literature in the behavioral and psychological sciences describing effect size indices and conversion formulas between different indices \citep[see e.g.][]{cohen_statistical_1988,borenstein_converting_2009,hedges_statistical_1985,ferguson_effect_2009,rosenthal_parametric_1994,long_regression_2006}.
\citet{cohen_statistical_1988} defined at least eight effect size indices defined for different models or types of dependent and independent variables and provided formulas to convert between the indices.
For example, Cohen's $d$ is defined for mean differences, $R^2$ is used for simple linear regression, and standardized log odds ratio is used in logistic regression.
Conversion formulas for these parametric indices are given in Table \ref{tab:conversions} and have been widely recognized and used in research and software \citep{cohen_statistical_1988,borenstein_converting_2009,lenhard_calculation_2017}.

Several authors have proposed robust effect size indices based on sample quantiles \citep{zhang_robust_1997,hedges_nonparametric_1984}.
These are robust in the sense that they do not assume a particular probability model, however, they are defined as specific parameters in the sense that they are a specific functional of the underlying distribution.

Despite the array of effect sizes, there are several limitations to the available indices:
1) there is no single unifying theory that links effect size indices.
2) as defined, many effect size indices do not accommodate nuisance covariates or multivariate outcomes and
3) each index is specific to a particular population parameter.
For example, Cohen's $d$ is designed for mean differences in the absence of covariates, existing semiparametric indices are quantile estimators, and correlation is specific to linear regression.
For these reasons, these classical effect size indices are not widely generalizable because their scale is dependent on the type of parameter.

In this paper, we define a new robust effect size index based on M-estimators.
M-estimators are parameter estimators that can be defined as the maximizer of an estimating equation.
This approach has several advantages over commonly used indices:
a) The generality of M-estimators makes the index widely applicable across many types of models that satisfy mild regularity conditions, including mean and quantile estimators, so this framework serves as a canonical unifying theory to link common indices.
b) The sandwich covariance estimate of M-estimators is consistent under model misspecification \citep{mackinnon_heteroskedasticity-consistent_1985,white_heteroskedasticity-consistent_1980}, so the index can accommodate unknown complex relationships between second moments of multiple dependent variables and the independent variable.
c) The robust effect size index is directly related to the Wald-style sandwich chi-squared statistic and is formulaically related to common indices.

Here, we describe sufficient conditions for the new effect size index to exist, describe how it relates to other indices, and show that other estimators can be biased under model misspecification.
In three examples, we show that the new index can be written as a function of Cohen's $d$, $R^2$, and a standardized log odds, demonstrating that it is related to indices that were developed using intuition for specific models.
In addition, we describe how to obtain a simple estimate of the index and provide functions to compute power or sample size given an effect size index and degrees of freedom of the target parameter.
Finally, we use simulations to assess the bias and variance of the proposed index estimator.

\section{Notation}
\label{sec:notation}
Unless otherwise noted, capital letters denote vectors or scalars and boldface letters denote matrices; lower and upper case greek letters denote vector and matrix parameters, respectively. Let $W_1=\{Y_1, X_1\}, \ldots, W_n=\{Y_n, X_n\}$ be a sample of independent observations from $\mathbb{W} \subset \R^p$ with associated probability measure $G$ and let $H$ denote the conditional distribution of $Y_i$ given $X_i$.
Here, $W_i$ denotes a combination of a potentially multivariate outcome vector $Y_i$ with a multivariate covariate vector $X_i$.

Let $W = \{W_1, \ldots, W_n\}$ denote the full dataset and $\theta^* \mapsto \Psi(\theta^*; W) \in \R$, $\theta^* \in \R^{m}$ be an estimating equation,
\begin{equation}\label{eq:mestimator}
\Psi(\theta^*; W) = n^{-1}\sum_{i=1}^n \psi(\theta^*; W_i),
\end{equation}
where $\psi$ is a known function. $\Psi$ is a scalar-valued function that can be maximized to obtain the M-estimator $\hat \theta$.
We define the parameter $\theta$ as the maximizer of the expected value of the estimating equation $\Psi$ under the true distribution $G$,
\begin{equation}\label{eq:thetaparameter}
\theta = \arg\max_{\theta^* \in \Theta} \E_G \Psi(\theta^* ; W) 
\end{equation}
and the estimator $\hat \theta$ is
\begin{equation*}
\hat\theta = \arg\max_{\theta^* \in \Theta}\Psi(\theta^* ; W).
\end{equation*}
Assume,
\begin{equation}
\label{eq:thetadef}
\theta = (\alpha, \beta),
\end{equation}
where $\alpha \in \R^{m_0}$ denotes a nuisance parameter, $\beta \in \R^{m_1}$ is the target parameter, and $m_0 + m_1 = m$.

We define the $m \times m$ matrices with $j,k$th elements
\begin{equation*}
\begin{aligned}
\mathbf{J}_{jk}(\theta) & = - \E_G\frac{\partial^2 \Psi(\theta^*; W)}{\partial\theta^*_{j} \partial\theta^*_{k} } \Big\vert_{\theta} \\
\mathbf{K}_{jk}(\theta) & = \E_G \frac{\partial \Psi(\theta^*; W)}{\partial\theta^*_{j} } \frac{\partial \Psi(\theta^*; W)}{ \partial\theta^*_{k} } \Big\vert_{\theta},
\end{aligned}
\end{equation*}
which are components of the asymptotic robust covariance matrix of $\sqrt{n}(\hat \theta - \theta)$.

\section{A new effect size index}
\subsection{Definition}
Here, we define a robust effect size that is based on the test statistic for
\begin{equation}
\label{eq:null}
H_0: \beta=\beta_0.
\end{equation}
$\beta_0$, is a reference value in the context of defining an effect size index.
Larger distances from $\beta_0$ represent larger effect sizes.
Under the regularity conditions in the Appendix,
\begin{equation}\label{eq:thetahatdist}
\sqrt{n}(\hat \theta - \theta) \sim N\left\{0, \mathbf{J}(\theta)^{-1} \mathbf{K}(\theta) \mathbf{J}(\theta)^{-1}\right\}.
\end{equation}
This implies that the typical robust Wald-style statistic for the test of \eqref{eq:null} is approximately chi-squared on $m_1$ degrees of freedom,
\begin{equation}\label{eq:teststatistic}
T_{m_1}(\hat\theta)^2 = n (\hat\beta - \beta_0)^T \Sigma_\beta(\hat\theta)^{-1} (\hat\beta-\beta_0) \sim \chi^2_{m_1}\left\{n (\beta-\beta_0)^T \Sigma_\beta(\theta)^{-1} (\beta - \beta_0) \right\},
\end{equation}
with noncentrality parameter $n (\beta-\beta_0)^T \Sigma_\beta(\theta)^{-1} (\beta - \beta_0)$, where $\Sigma_\beta(\theta)$ is the asymptotic covariance matrix of $\hat\beta$ is derived from the covariance of \eqref{eq:thetahatdist} \citep{boos_essential_2013,van_der_vaart_asymptotic_2000}.
We define the square of the effect size index as the component of the chi-squared statistic that is due to the deviation of $\beta$ from the null value:
\begin{equation}\label{eq:ses}
S_\beta(\theta)^2 = (\beta-\beta_0)^T \Sigma_\beta(\theta)^{-1} (\beta - \beta_0).
\end{equation}
As we demonstrate below, the covariance $\Sigma_\beta(\theta)$ serves to standardize the parameter $\beta$ so that it is unitless.
The regularity conditions given in the appendix are sufficient for the index to exist.
The robust index, $S_\beta(\theta) = \sqrt{S_\beta(\theta)^2}$, is defined as the square root of $S_\beta(\theta)^2$ so that the scale is proportional to that used for Cohen's $d$ (see Example \ref{ex:eeexample}).

This index has several advantages: it is widely applicable because it is constructed from M-estimators; it relies on a robust covariance estimate; it is directly related to the robust chi-squared statistic; it is related to classical indices, and it induces several classical transformation formulas \citep{cohen_statistical_1988,borenstein_converting_2009,lenhard_calculation_2017}.

\subsection{An estimator}
$S_\beta(\theta)$ is defined in terms of parameter values and so must be estimated from data when reported in a study.
Let $T_{m_1}(\hat\theta)^2$ be as defined in \eqref{eq:teststatistic}, then
\begin{equation}\label{eq:estimate}
\hat S_\beta(\theta) = \left\{\max\left[0, (T_{m_1}(\hat\theta)^2 - m)/(n-m)\right]\right\}^{1/2}
\end{equation}
is consistent for $S_\beta(\theta)$, which follows by the consistency of the components that make up $T_{m_1}(\hat\theta)^2$ \citep{van_der_vaart_asymptotic_2000,white_heteroskedasticity-consistent_1980}.
We use the factor $(n-m)$ to account for the estimation of $m$ parameters.

There is considerable existing research on estimators for noncentrality parameters of chi-squared statistics \citep{saxena_estimation_1982,chow_complete_1987,neff_further_1976,kubokawa_estimation_1993,shao_improving_1995,lopez-blazquez_unbiased_2000}.
While the estimator \eqref{eq:estimate} is inadmissable  \citep{chow_complete_1987}, it has smaller risk than the usual unbiased maximum likelihood estimator, $S^2 = (T_{m_1}(\hat\theta)^2 - m)/(n-m)$, and is easy to compute.
We assess estimator bias in Section \ref{sec:simulations}.

\section{Examples}
\label{sec:examples}
In this section we show that this robust index yields several classical effect size indices when the models are correctly specified.
We demonstrate the interpretability of the effect size index through a series of examples.
The following example shows that the robust index for a difference in means is equal to Cohen's $d$, provided that the parametric model is correctly specified.

\begin{example}[Difference in means]\label{ex:eeexample}
 In this example we consider a two mean model, where $W_i = \{Y_i, X_i\}$ and the conditional mean of $Y_i$ given $X_i$ converges. That is,
\begin{equation}\label{eq:exmean}
n_x^{-1} \sum_{i : X_i=x}^{n_x} \E( Y_{i} \mid X_i=x ) \xrightarrow{p} \mu_x \in \R,
\end{equation}
for independent observations $i=1,\ldots, n$, where $x,X_i \in \{0,1\}$, $n_x = \sum_{i=1}^n I(X_i=x)$, and we assume the limit \eqref{eq:exmean} exists.
In addition we assume $\P(X_i =1)=\pi_1 = 1-\pi_0$ is known and
\[
n_x^{-1} \sum_{i : X_i=x}^{n_x}\text{Var}(Y_i \mid X_i=x) \xrightarrow{p} \sigma^2_x < \infty.
\]
Let $\partial\Psi(\theta; W)/\partial \theta = n^{-1}\sum_{i=1}^n \{(2X_i - 1)\pi_{X_i}^{-1}Y_i - \theta \}$, then
\begin{align*}
\hat \theta & = \frac{n_1}{n} \pi_1^{-1}\hat\mu_1 - \frac{n_0}{n} \pi_0^{-1}\hat\mu_0\\
\E \hat \theta & = \mu_1 - \mu_0 \\
J(\theta) & = 1 \\
K(\theta) & = \lim_{n\to\infty}  n^{-1}\sum_{i,j} \E_H \left\{ (2 X_i - 1)\pi_{X_i}^{-1}Y_i - \theta\right\}\left\{(2 X_j - 1) \pi_{X_j}^{-1}Y_j -  \theta\right\},
\end{align*}
where $\hat \mu_x = n_x \sum_{i : X_i=x}^{n_x} Y_i$.
If the mean model is correctly specified, as in \eqref{eq:exmean}, then $K(\theta) = \lim_{n\to\infty} n^{-1}\sum_{i=1}^n \pi_{X_i}^{-2}\text{Var}(Y_i \mid X_i)$.
Note that $\Psi$ in this example is not defined as the derivative of a log-likelihood:
it defines a single parameter that is a difference in means and does not require each observation to have the same distribution.
This approach still allows us to determine the asymptotic variance of $n^{1/2}\hat\theta$,
\begin{align*}
J(\theta) K(\theta)^{-1}J(\theta)
& = \lim_{n\to\infty} n^{-1}\sum_{i=1}^n \pi_{X_i}^{-2}\text{Var}(Y_i \mid X_i) \\
& = \lim_{n\to\infty} n^{-1}\left\{n_1\pi_{1}^{-2}\sigma^2_1 + n_0\pi_{0}^{-2}\sigma^2_0 \right\} \\
& = \pi_1^{-1}\sigma^2_1 + \pi_0^{-1}\sigma^2_0.
\end{align*}
Then the robust effect size \eqref{eq:ses} is
\begin{equation}\label{eq:sesCohen}
S_\beta(\theta) = \sqrt{\frac{(\mu_1 - \mu_0)^2}{\pi_1^{-1}\sigma^2_1 + \pi_0^{-1}\sigma^2_0}}.
\end{equation}
For fixed sample proportions $\pi_0$ and $\pi_1$, when $\sigma_0^2 = \sigma_1^2$, $S_\beta(\theta)$ is proportional to the square of the classical index of effect size for the comparison of two means, Cohen's $d$ \citep{cohen_statistical_1988}.
However, $S_\beta(\theta)$ is more flexible: it can accommodate unequal variance among groups and accounts for the effect that unequal sample proportions has on the power of the test statistic.
Thus, $S$ is an index that accounts for all features of the study design that will affect the power to detect a difference.
In this example, we did not explicitly assume a distribution for $X_i$, only that the variance of $Y_i$ given $X_i$ converges in probability to a constant.
\end{example}

The following example derives the robust effect size for simple linear regression.
This is the continuous independent variable version of Cohen's $d$ and is related to $R^2$.

\begin{example}[Simple linear regression] \label{ex:SLRexample}
Consider the simple linear regression model
\[
Y_i = \alpha + X_i \beta + \epsilon_i
\]
where $\alpha$ and $\beta$ are unknown parameters, $Y_i \in \R$, $X_i \in \R$ and $\epsilon_i$ follows an unknown distribution with zero mean and conditional variance that can depend on $X_i$, $\text{Var}(Y_i \mid X_i) = \sigma^2(X_i)$.
Let $\Psi(\theta; W_i) = n^{-1} \sum_{i=1}^n(Y_i - \alpha - X_i \beta)^2/2$.
In this model
\begin{equation}\label{eq:SLRmat}
\begin{aligned}
\mathbf{J}(\theta)^{-1} &=
\sigma^{-2}_x\begin{bmatrix}
\sigma^2_x + \mu^2_x & - \mu_x \\
- \mu_x & 1 \\
\end{bmatrix} \\
\mathbf{K}(\theta) & =
\begin{bmatrix}
\sigma^2 & \mu_{xy} \\
\mu_{xy} & \sigma^2_{xy} +2 \mu_x \mu_{xy} - \mu_{x}^2\sigma^2 \\
\end{bmatrix}
\end{aligned}
\end{equation}
where
\begin{equation}\label{eq:SLRmatcomps}
\begin{aligned}
\mu_x & = \E_G  X_i\\
\sigma^2_x & = \E_G  (X_i - \mu_x)^2\\
\sigma^2 & = \E_G  (Y_i-\alpha-X_i\beta)^2\\
\mu_{xy} & = \E_G  X_i(Y_i-\alpha-X_i\beta)^2\\
\sigma^2_{xy} & = \E_G  (X_i - \mu_x)^2(Y_i-\alpha-X_i\beta)^2.
\end{aligned}
\end{equation}
After some algebra, combining the formulas \eqref{eq:SLRmat} and \eqref{eq:SLRmatcomps} gives
\begin{align*}
\Sigma_\beta & = \sigma^{-4}_x\sigma^2_{xy}.
\end{align*}
Then \eqref{eq:ses} is
\begin{equation}\label{eq:SLRses}
S_\beta(\theta)^2 = \frac{\sigma_x^4}{\sigma^2_{xy}}\beta^2.
\end{equation}
The intuition of \eqref{eq:SLRses} is best understood by considering the homoskedastic case where $\E_H (Y_i-\alpha-X_i\beta)^2 = \sigma^2$ for all $i=1,\ldots, n$.
Then, $\sigma_x^4/\sigma^2_{xy} \beta^2 = \sigma_x^2/\sigma^2 \beta^2$.
This is similar to $R^2$, except that the denominator is the variance of $Y_i$ conditional on $X_i$ instead of the marginal variance of $Y_i$.
\end{example}

In the following example we introduce two levels of complexity by considering logistic regression with multidimensional nuisance and target parameters.
\begin{example}[Logistic regression with covariates] \label{ex:LR}
For logistic regression we utilize the model
\begin{equation}
\label{eq:LRmodel}
\E (Y_i \mid X_i) = \text{expit}(X_{i0} \alpha + X_{i1} \beta) = \text{expit}(X_i \theta),
\end{equation}
where $Y_i$ is a Bernoulli random variable, $X_i = [X_{i0}, X_{i1}] \in \R^{p-1}$ is a row vector, and $\alpha$ and $\beta$ are as defined in \eqref{eq:thetadef}.
Let $\mathbf{X} = [ X_1^T \ldots X_n^T]^T \in \R^{n \times (p-1)}$ and similarly define $\mathbf{X}_0$ and $\mathbf{X}_1$.
Let $\mathbf{P} \in \R^{n \times n}$ be the matrix with $\mathbf{P}_{ii} = \text{expit}(X_i \theta)\left\{1-\text{expit}(X_i \theta)\right\}$ and $\mathbf{P}_{ij} = 0$ for $i\ne j$.
Let $\mathbf{Q} \in \R^{n \times n}$ be the matrix with $\mathbf{Q}_{ii} = \{ Y_i - \text{expit}(X_i \theta)\}^2$  and $\mathbf{Q}_{ij} = 0$ for $i\ne j$.
If \eqref{eq:LRmodel} is correctly specified then $\E_{H}(\mathbf{P}_{ii} \mid X_i)  = \E_{H}(\mathbf{Q}_{ii} \mid X_i) = \text{Var}(Y_i \mid X_i)$.
If this equality does not hold then there is under or over dispersion.

To find the robust effect size, we first need to find the covariance matrix of $\hat\beta$.
To simplify notation we define the matrices
\[
\mathbf{A}_{k\ell}(\mathbf{P}) = \E_G n^{-1}\mathbf{X}_k^T \mathbf{P} \mathbf{X}_\ell
\]
for $k,\ell = 0,1$.
The block matrix of $\mathbf{J}_G(\theta)^{-1}$ corresponding to the parameter $\beta$ is
\begin{equation}\label{eq:LRfisherinfo}
\mathbf{I}_\beta(\theta)^{-1} = \left\{\mathbf{A}_{11}(\mathbf{P}) - \mathbf{A}_{10}(\mathbf{P}) \mathbf{A}_{00}(\mathbf{P})^{-1} \mathbf{A}_{01}(\mathbf{P})\right\}^{-1}.
\end{equation}
Equation \eqref{eq:LRfisherinfo} is the asymptotic covariance of $\hat \beta$, controlling for $\mathbf{X}_0$, if model \eqref{eq:LRmodel} is correctly specified.

The robust covariance for $\beta$ can be derived by finding the block matrix of $J(\theta)^{-1} K(\theta) J(\theta)^{-1}$ corresponding to $\beta$.
In this general case, the asymptotic covariance matrix of $\hat\beta$ is
\begin{align*}
\Sigma_\beta(\theta) = & \mathbf{I}_\beta(\theta)^{-1}  \left[   \mathbf{A}_{10}(\mathbf{P}) \mathbf{A}_{00}(\mathbf{P})^{-1} \mathbf{A}_{00}(\mathbf{Q}) \mathbf{A}_{00}(\mathbf{P})^{-1} \mathbf{A}_{01}(\mathbf{P}) \right. \\
 & \left. - \mathbf{A}_{10}(\mathbf{P})\mathbf{A}_{00}(\mathbf{P})^{-1} \mathbf{A}_{01}(\mathbf{Q}) \right] \mathbf{I}_\beta(\theta)^{-1} \\
& + \mathbf{I}_\beta(\theta)^{-1}  \left [   \mathbf{A}_{11}(\mathbf{Q}) - \mathbf{A}_{10}(\mathbf{Q})\mathbf{A}_{00}(\mathbf{P})^{-1} \mathbf{A}_{01}(\mathbf{P})   \right]  \mathbf{I}_\beta(\theta)^{-1}.
\end{align*}
If the model is correctly specified, $\mathbf{P} = \mathbf{Q}$, $\Sigma_\beta(\theta) = \mathbf{I}_\beta(\theta)^{-1}$, then
\begin{equation}\label{eq:LRses}
S_\beta(\theta) = \sqrt{\beta^T \mathbf{I}_\beta(\theta) \beta}.
\end{equation}
The parameter \eqref{eq:LRses} describes the effect of $\beta$ controlling for the collinearity of variables of interest $\mathbf{X}_1$, with the nuisance variables, $\mathbf{X}_0$.
If the collinearity is high, then the diagonal of $\mathbf{I}_\beta(\theta)^{-1}$ will be large and the effect size will be reduced.

Many suggestions have been made to compute standardized coefficients in the context of logistic regression \citep[for a review see][]{menard_six_2004,menard_standards_2011}.
The square of the robust index in this context, under correct model specification, is the square of a fully standardized coefficient and differs by a factor of $\sqrt{n}$ from the earliest proposed standardized index \citep{goodman_modified_1972}.
The index proposed by \citet{goodman_modified_1972} is simply a wald statistic and was rightly criticized for its dependence on the sample size \citep{menard_standards_2011}, despite that it correctly accounts for the fact that the variance of a binomial random variable is a function of its mean through the use of the diagonal matrix $\mathbf{P}$ in the matrix $\mathbf{I}_\beta(\theta)$.
The robust index remediates the dependence that Goodman's standardized coefficient has on the sample size.

\end{example}

%

\section{Relation to other indices}

The robust index can be expressed as a function of several common effect size indices for continuous or dichotomous dependent variables when there is homoskedasticity (Figure \ref{fig:conversioncurves}; Table \ref{tab:conversions}).
The relations between effect sizes implied by the new index yields the classical conversion formulas between effect sizes  \citep{borenstein_converting_2009,selya_practical_2012}.
While the index is related to existing indices under correct model specification, the advantage of the robust index is that it is defined if the variance model is incorrectly specified.
This is the case, for example, in linear regression when there is heteroskedasticity and the model assumes a single variance term for all subjects or in logistic regression when there is over dispersion.
By using the formulas in Table \ref{tab:conversions}, we can obtain robust versions of classical indices by writing them as a function of $S_\beta$

\begin{knitrout}
\definecolor{shadecolor}{rgb}{0.969, 0.969, 0.969}\color{fgcolor}\begin{figure}

{\centering \includegraphics[width=\maxwidth]{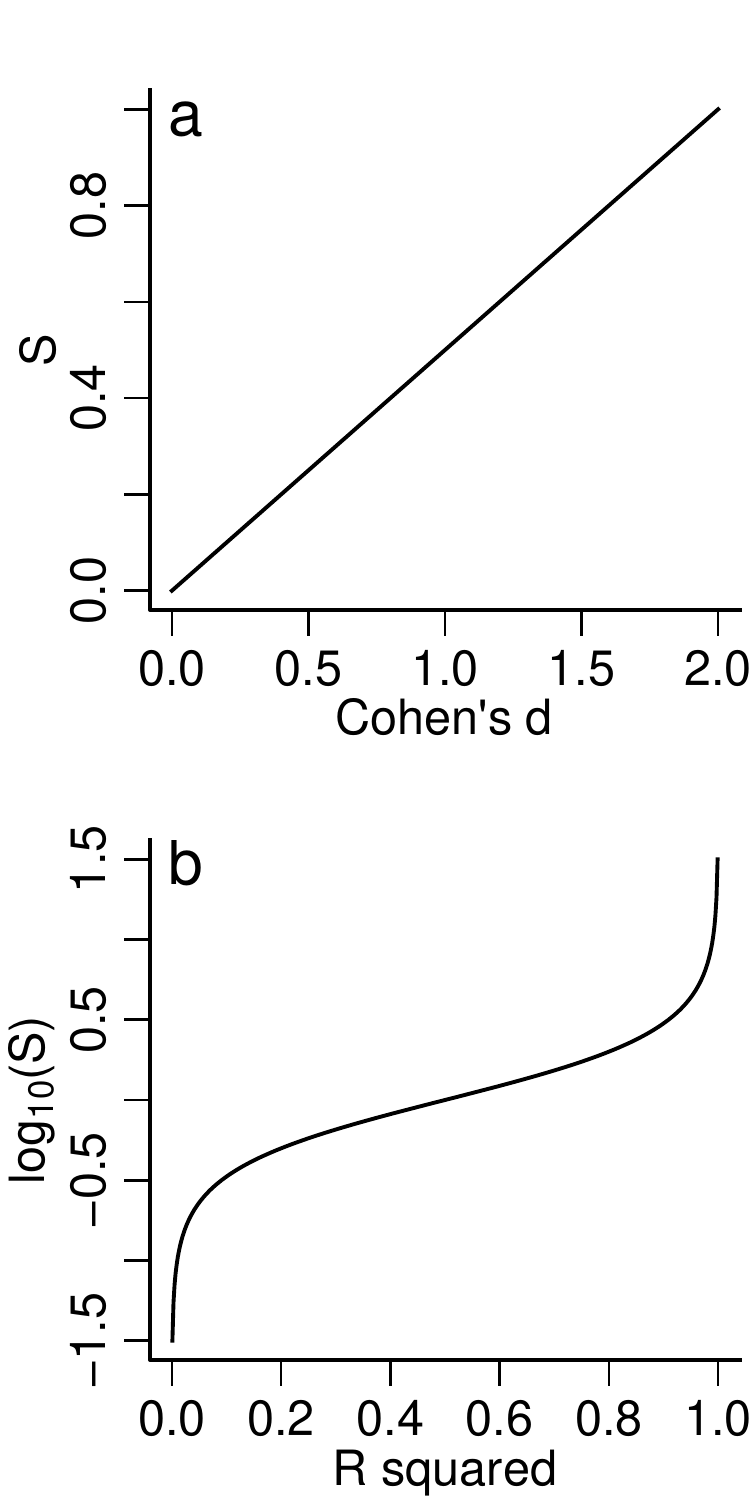}

}

\caption{Graphs of the robust effect size as a function of some common effect size indices (see formulas in Table \ref{tab:conversions}. (a) Cohen's $d$, when $\pi_0=\pi_1=1/2$ and $\sigma_0 = \sigma_1$; (b) $R^2$.}\label{fig:conversioncurves}
\end{figure}

\end{knitrout}

\citet{cohen_statistical_1988} defined ranges of meaningful effect sizes for the behavioral sciences (Table \ref{tab:essuggestions}).
These intervals can also be used to define similar regions for the robust index.
These recommendations serve as a useful guide, however, ranges of meaningful effect sizes are field specific and should be based on clinical expertise and the effect an intervention could have if applied to the population of interest.

\renewcommand{\arraystretch}{1.3}
\begin{table}
\begin{tabular}{c|cccc}
$\mapsto$ & $\lvert d \rvert$ & $f^2_\beta$ & $R^2_\beta$ & $S_\beta$ \\\hline
$d$ & $\lvert d \rvert$ & $(\pi_1^{-1} + \pi_0^{-1})^{-1}\times d^2$ & $\frac{d^2}{(\pi_1^{-1} + \pi_0^{-1}) + d^2}$ & $(\pi_1^{-1} + \pi_0^{-1})^{-1/2}\times\lvert d \rvert $ \\
$f^2_\beta$ & $(\pi_1^{-1} + \pi_0^{-1})^{1/2}\times \sqrt{f^2_\beta} $& $f^2_\beta$ & $\frac{f_\beta^2}{1+ f^2}$  & $\sqrt{f^2_\beta}$ \\
$R^2_\beta$ &$ (\pi_1^{-1} + \pi_0^{-1})^{1/2}\times \sqrt{\frac{R^2_\beta}{1-R^2}}$  & $\frac{R^2_\beta}{1-R^2}$& $R^2_\beta$ & $\sqrt{\frac{R^2_\beta}{1-R^2}}$ \\
$S_\beta$ & $(\pi_1^{-1} + \pi_0^{-1})^{1/2}\times S_\beta $ & $S^2_\beta$ & $\frac{S_\beta^2}{1+ S^2}$& $S_\beta$

\end{tabular}
\caption{Effect size conversion formulas based on derivations from the robust index under homoskedasticity. Each row denotes the input argument and the column denotes the desired output value. Robust versions of classical values can be obtained by computing them as a function of $S_\beta$. $\pi_1$ and $\pi_0$ denote the population proportions of each group for a two sample comparison. $d$ is Cohen's $d$, $f^2_\beta$ is Cohen's effect size for multiple regression, $R^2_\beta$ is the partial coefficient of determination, $S_\beta$ is the robust index.
The variables without subscripts denote the value for the whole model.
Conversion formulas derived by the robust index match classical formulas \citep{cohen_statistical_1988,borenstein_converting_2009,lenhard_calculation_2017}.}
\label{tab:conversions}
\end{table}

\renewcommand{\arraystretch}{1}
\begin{table}
\begin{center}
\begin{tabular}{c|cc}
Effect size & $d$ & $S$\\\hline
None-Small & $[0, 0.2]$ & $[0, 0.1]$ \\
Small-Medium & $(0.2, 0.5]$ & $(0.1, 0.25]$ \\
Medium-Large & $(0.5, 0.8]$ & $(0.25, 0.4]$
\end{tabular}
\end{center}

\caption{Effect size thresholds suggested by \cite{cohen_statistical_1988} on the scale of $d$ and the robust index ($S_\beta$), using the formula from Table \ref{tab:conversions} assuming equal sample proportions.}
\label{tab:essuggestions}
\end{table}

\subsection{Bias of existing indices under model misspecification}

To understand the bias of the classical estimators under model misspecification, we compare the asymptotic value of the classical estimators to the effect size formulas in Table \eqref{tab:conversions}.
Under model misspecification, the existing parametric effect size indices can be biased.

The estimator for Cohen's $d$ using pooled variance converges to
\[
\hat d_C = \frac{\hat \mu_1 - \hat \mu_0}{\frac{(n_1-1)\hat\sigma_1^2 + (n_0-1)\hat\sigma_0^2}{n_1 + n_0 -2} } \to_p \frac{\mu_1-\mu_0}{\pi_1 \sigma^2_1 + (1-\pi_1)\sigma^2_0} = d_C.
\]
Taking the ratio of this value to the robust value of Cohen's $d$ in Table \ref{tab:conversions} gives
\[
d_C/d(S) = (\pi_1^{-1} + (1-\pi_1)^{-1})^{-1/2} \times \left(\frac{\pi_1^{-1} \sigma^2_1 + (1-\pi_1)^{-1} \sigma^2_0}{\pi_1 \sigma^2_1 + (1-\pi_1) \sigma^2_0}\right)^{1/2}
\]
A plot of this ratio with respect to $\log_2(\sigma^2_1/\sigma^2_0)$ and $\pi_1$ is given in Figure \ref{fig:cohensdbias}.
When $\pi_1 = 1/2$ or $\sigma^2_1 = \sigma^2_0$ then there is no bias.
When $\pi_1<1/2$ and $\sigma^2_1>\sigma^2_0$ Cohen's $d$ overestimates the effect size.
When $\pi_1<1/2$ is small and $ \sigma^2_1<\sigma^2_0$ Cohen's $d$ under underestimates the effect size.
The plot is symmetric about the point $(0,1/2)$.

The classical estimator for $R^2$ converges to
\[
R^2_C = \frac{\sigma^2_x \beta^2}{\sigma^2_x \beta^2 + \sigma^2_y}.
\]
Taking the ratio of this value and the formula for $R^2(S)$ given in Table \ref{tab:conversions} gives,
\[
R^2_C/R^2(S) = \frac{\sigma^4_x \beta^2 + \sigma^2_x\sigma^2_y}{\sigma^4_x \beta^2 + \sigma^2_{xy}},
\]
where variables are as defined in \eqref{eq:SLRmatcomps}.
Figure \ref{fig:cohensdbias} plots the bias as a function of $\log_2\{\sigma^2_{xy}/(\sigma^2_x\sigma^2_y)\}$.
When the variance is constant across subjects, $\mathrm{Var}(Y_i \mid X_i)  = \sigma^2_y$, then the bias is zero.
If not, then the direction of the bias of the classical estimator depends on the relationship between $\mathrm{Var}(Y_i \mid X_i)$ and $X_i$.
\begin{knitrout}
\definecolor{shadecolor}{rgb}{0.969, 0.969, 0.969}\color{fgcolor}\begin{figure}

{\centering \includegraphics[width=\maxwidth]{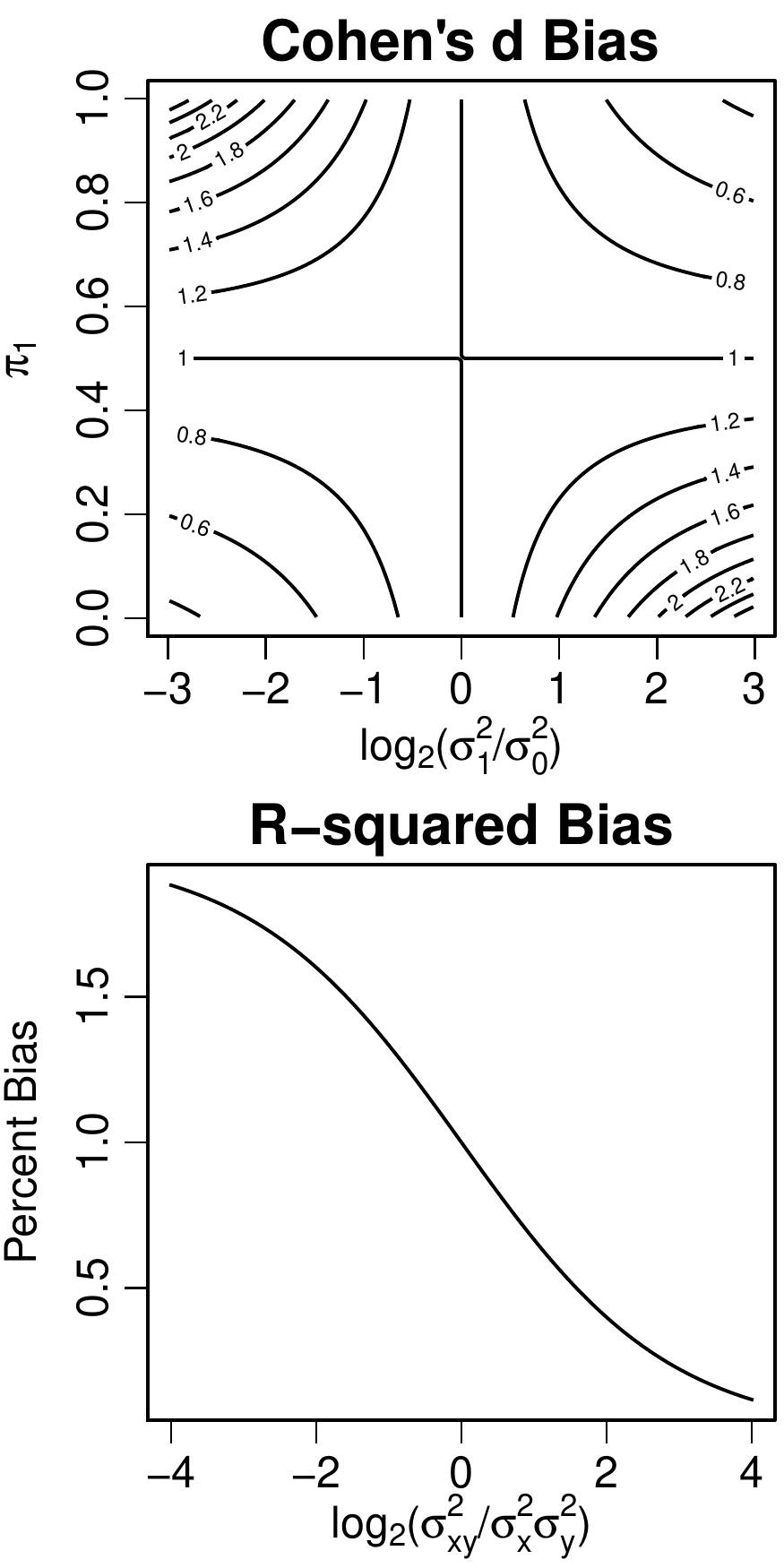}

}

\caption{Percent bias for Cohen's $d$ and $R^2$. When $\pi_1=1/2$ or the variances are equal the classical estimator of Cohen's $d$ is unbiased, however it can be positively or negatively biased when the variances and sampling proportions are not equal. Similarly for $R^2$, when $\text{Var}(Y_i \mid X_i)$ is constant across subjects, there is no bias (because $\sigma^2_{xy}=\sigma^2_x \sigma^2_y$), but when this is not true, the classical estimator can be positively or negatively biased depending on the relationship between the variances. Variables are as defined in \eqref{eq:SLRmatcomps}}\label{fig:cohensdbias}
\end{figure}

\end{knitrout}

\section{Determining effect sizes, sample sizes, and power}

A convenient aspect of the robust index is that it makes power analyses easier.
The formula is the same for every parameter that is a solution to an estimating equation such as \eqref{eq:thetaparameter}.
For a fixed sample size and rejection threshold, power can be determined from the robust index and degrees of freedom of the chi-squared test using \eqref{eq:teststatistic}.
The explicit formula for power can be written
\begin{equation}\label{eq:power}
1- t_2 = 1-\Phi_\text{df}\left\{ \Phi^{-1}_\text{df}(1- t_1; 0); n\times S_\beta(\theta)^2 \right\},
\end{equation}
where $t_1$ and $t_2$ denote the type 1 and type 2 error rates, respectively, df denotes the degrees of freedom of the test statistic, $\Phi(\cdot; \lambda)$ denotes the cumulative distribution function of a noncentral chi-squared distribution with noncentrality parameter $\lambda$, and $S_\beta$ is as defined in \eqref{eq:ses}.
Equation \eqref{eq:power} can be easily solved for sample size, power, error rate, or effect size, using basic statistical software with fixed values of the other variables (Figure \ref{fig:powercurves}).
Because the robust index is not model dependent, power curves are effectively model-free and applicable for any fixed sample size, rejection threshold, and degrees of freedom.

\begin{knitrout}
\definecolor{shadecolor}{rgb}{0.969, 0.969, 0.969}\color{fgcolor}\begin{figure}

{\centering \includegraphics[width=\maxwidth]{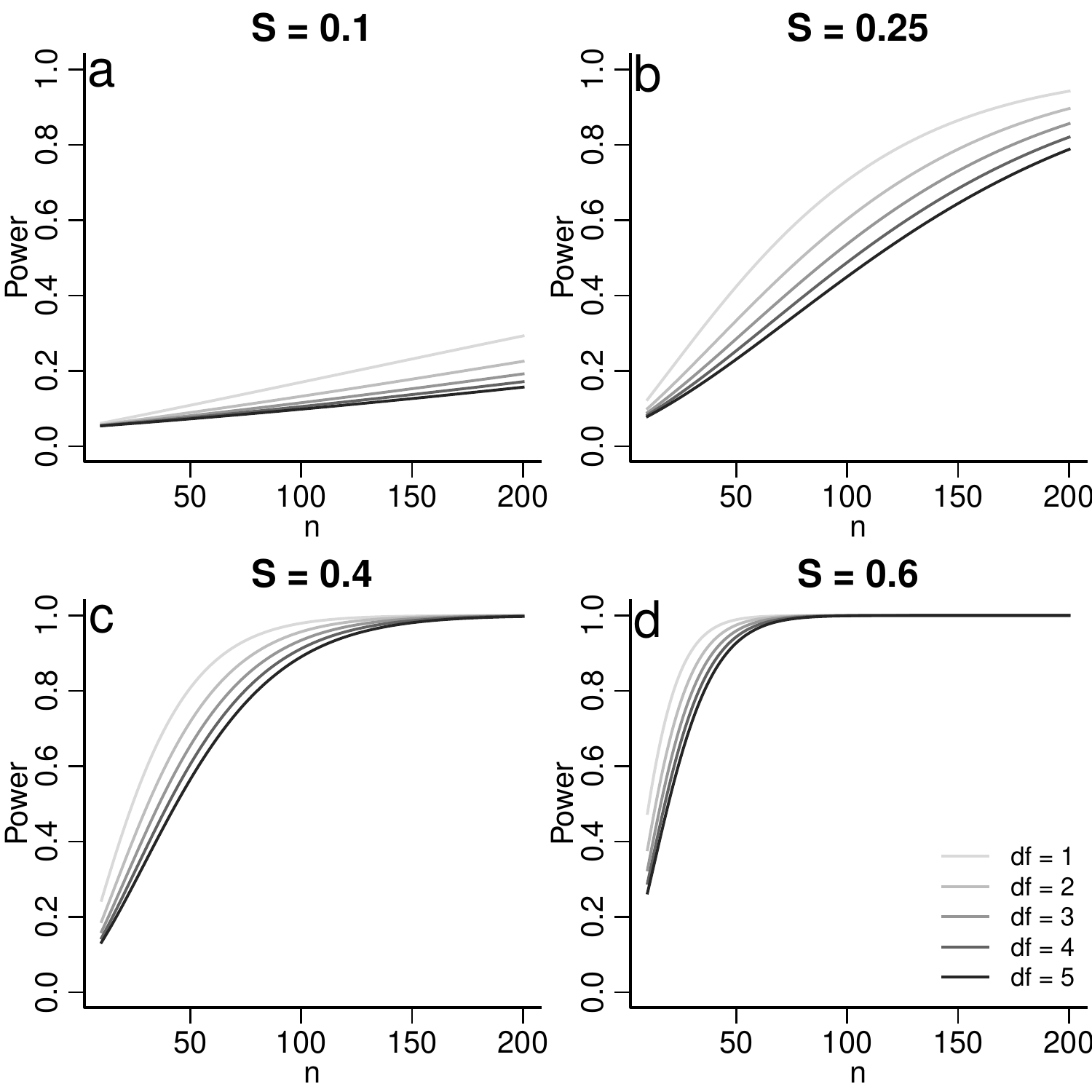}

}

\caption{Power curves as a function of the sample size for several values of the robust index ($S$) and degrees of freedom (df), for a rejection threshold of $\alpha=0.05$. The curves are given by formula \eqref{eq:power} and are not model dependent.}\label{fig:powercurves}
\end{figure}

\end{knitrout}

\section{Simulation analysis}
\label{sec:simulations}
We used 1,000 simulations to assess finite sample bias.
Covariates of row vectors, $X_i$, were generated from a multivariate normal distribution
$X_i \sim N(0, \Sigma_X)$,
where,
\[
\Sigma_X
=
\begin{bmatrix}
I_{m_0} & \rho^2/(m_0 m_1) \times \boldsymbol{1}_{m_0}\boldsymbol{1}_{m_1}^T\\
\rho^2/(m_0 m_1) \times \boldsymbol{1}_{m_1}\boldsymbol{1}_{m_0}^T& I_{m_1}
\end{bmatrix}
\]
with $\rho^2 \in \{0, 0.6\}$, $m_0 \in \{2, 5\}$, and $m_1 \in \{1,3,5\}$. Here, $I_{m_0}$ and $\boldsymbol{1}_{m_0}$ denote the $m_0\times m_0$ identity matrix and a vector of ones in $\R^{m_0}$, respectively.
This distribution implies that the total correlation between the nuisance covariates and target covariates is equal to $\rho^2$.
Samples of $Y_i$, for $i=1, \ldots, n$ of size $n \in \{25, 50, 100, 250, 500, 1000\}$ were generated with mean
\[
\E Y_i = \beta X_{i1}\boldsymbol{1}_{m_1},
\]
where $\beta$ was determined such that $S \in \{0, 0.1, 0.25, 0.4, 0.6 \}$.
We used a gamma distribution with shape parameter $a\in\{0.5, 10 \}$ and rate equal to $\sqrt{a/ X^2_{i,m_0+1}} $ to generate the errors for $Y_i$.
For each simulation, we compute bias of the estimator \eqref{eq:estimate}.
Only a subset of the results are reported here, however code to run the simulations and the saved simulation results are published with this paper.


Bias and variance of the estimator is given presented for $\rho^2\in\{0, 0.6\}$ and all values of $S$ considered in the simulations for $m_0=2$ (Figure \ref{fig:simres}).
Results demonstrate the the effect size estimator is biased upwards in small samples, but the bias is close to zero for sample sizes over 500. Because the effect size is defined conditional on covariates, the existence of covariates does not affect estimation bias.
The standard error of the estimator is larger in small samples and for larger values of $S$.
When the sample size is small, $n=25$, the standard error is nearly equal to the value of $S$.

\begin{knitrout}
\definecolor{shadecolor}{rgb}{0.969, 0.969, 0.969}\color{fgcolor}\begin{figure}

{\centering \includegraphics[width=\maxwidth]{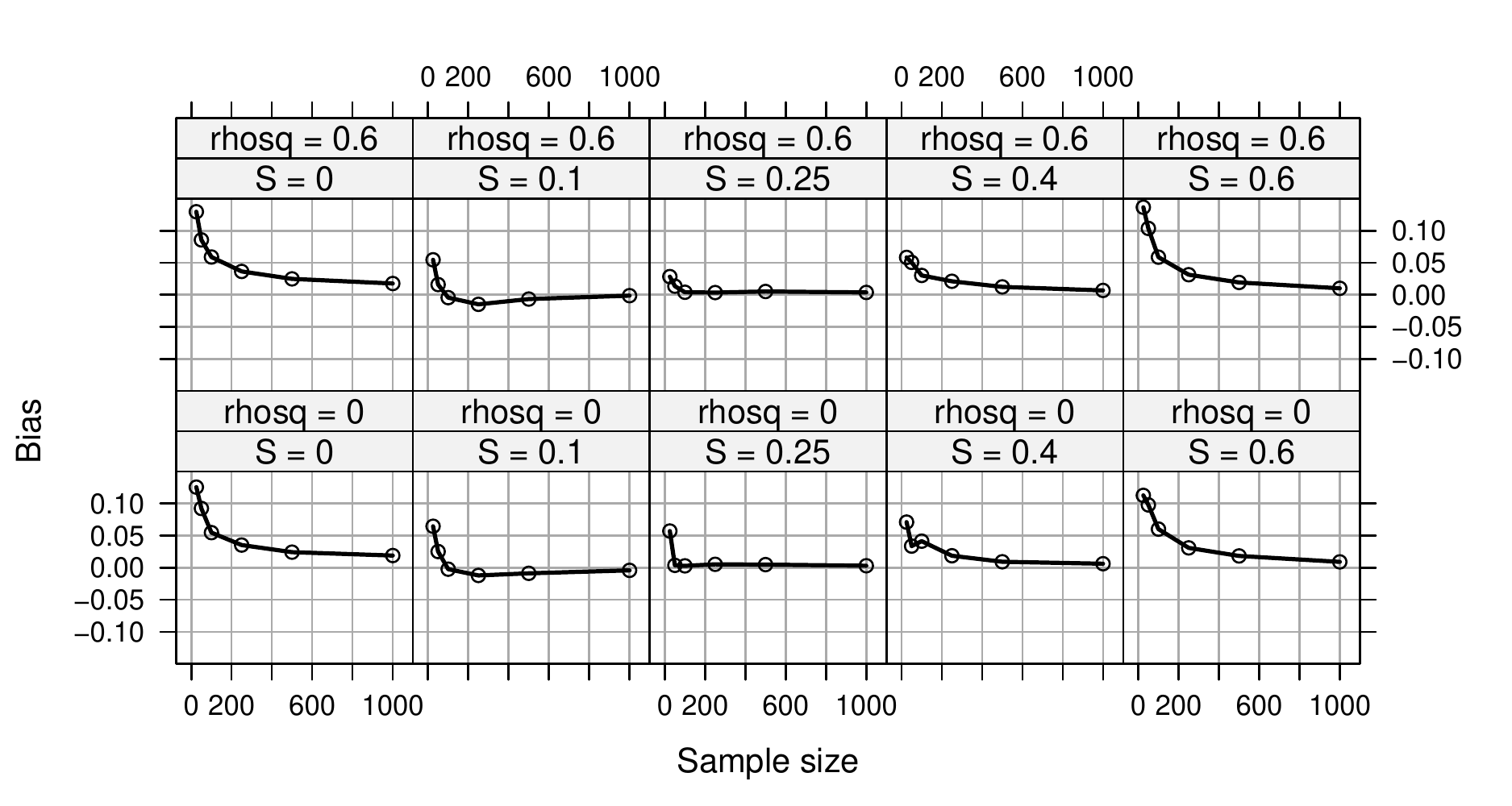}
\includegraphics[width=\maxwidth]{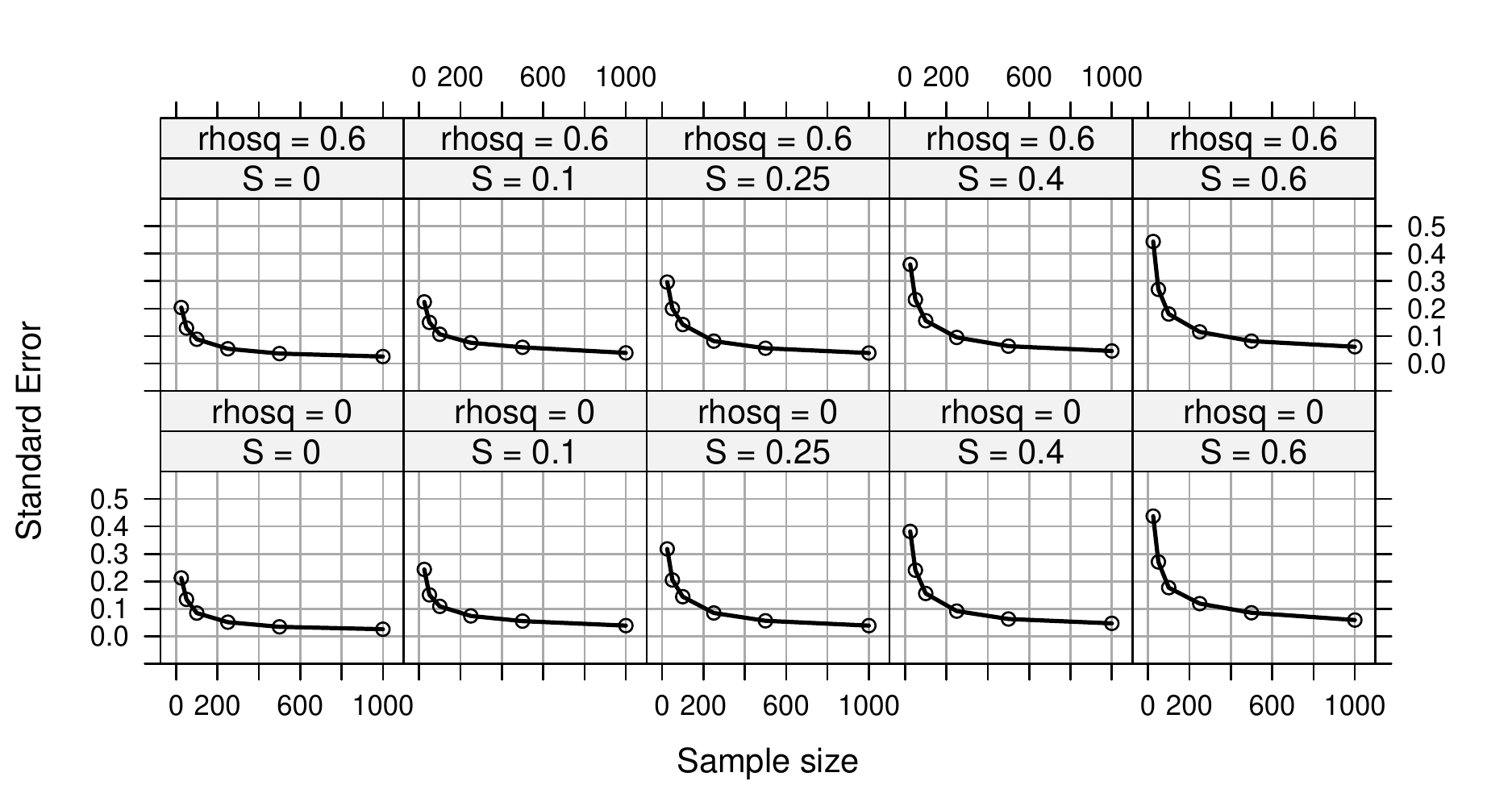}

}

\caption[Bias and variance of $\hat S$ when the data generating distribution has skew$=$0.63 with two nuisance covariate ($m_0=2$)]{Bias and variance of $\hat S$ when the data generating distribution has skew$=$0.63 with two nuisance covariate ($m_0=2$). $\hat S$ tends to be positively biased across values of $S$. Th standard error is proportional to $S$ and is quite large in small samples. Rhosq denotes the total squared correlation of nuisance covariates with the target variables. Rhosq does not appear to affect the bias, variance, or value of the effect size index because it is defined conditionally on the covariates.}\label{fig:simres}
\end{figure}

\end{knitrout}


\section{Discussion}
We proposed a robust effect size index that utilizes an M-estimator framework to define an index that is generalizable across a wide range of models.
The robust index provides a unifying framework for formulaically relating effect sizes across different models.
The proposed index is robust to model misspecification, easy to estimate, and related to classical effect size indices.
We showed that classical estimators can be negatively or positively biased when the covariance model is misspecified.

The relationship between the robust index and indices based on correctly specified models (such as Cohen's $d$ and $R^2$) is appealing because it follows intuition from other areas of robust covariance estimation. That is, when the estimating equation is proportional to the log likelihood, then the robust index is a function of classical definitions derived from likelihood based models.
The new framework also generalizes classical indices by easily accommodating nuisance covariates and sandwich covariance estimators that are robust to heteroskedasticity.
The robust index puts indices for all models on the same scale so that asymptotically accurate power analyses can be performed for model parameters using a single framework.

One important feature of the proposed index is that it is defined conditional on covariates.
While the effect size lies on a standardized scale that is related directly to the power of the test, the inclusion of covariates affects the interpretation of the index because it is defined conditional on the covariates.
For this reason, careful consideration of the target parameter is necessary for accurate interpretation and comparison across studies that present the robust index.
Marginal estimators (without conditioning on covariates) should be considered if the investigator is interested in the general effect across a given population.

Several limitations may inspire future research topics:
like p-values, estimates of effect size indices can be subject to bias by data dredging.
Moreover, the motivation for the index is based on asymptotic results, and can be inaccurate for small samples.
Thus, methods for bias adjustment or low mean squared error estimators could be considered to adjust the effects of data dredging or small sample sizes.
Here, we considered an M-estimator framework, but a semiparametric or robust likelihood framework may have useful properties as well \citep{royall_interpreting_2003,blume_statistical_2007}.
For this reason, we believe this index serves as a first step in constructing a class of general robust effect size estimators that can make communication of effect sizes uniform across models in the behavioral sciences.

\section*{Funding}
National Institutes of Health grants: (5P30CA068485-22 to S.N.V.).

\section{Appendix}

The following regularity conditions are required for the asymptotic normality of $\sqrt{n}(\hat\theta - \theta)$  \citep{van_der_vaart_asymptotic_2000}
\begin{enumerate}[a)]
\item The function $\theta^* \mapsto \Psi(\theta^*; w)$ is almost surely differentiable at $\theta$ with respect to $G$, where objects are as defined in \eqref{eq:mestimator} and \eqref{eq:thetaparameter}.
\item For every $\theta_1$ and $\theta_2$ in a neighborhood of $\theta$ and measurable function $m(w)$ such that $\E_G m(W)^2 < \infty$, $\lvert \Psi(\theta_1;w) - \Psi(\theta_2;w) \rvert \le m(w) \lVert \theta_1 - \theta_2\rVert$.
\item The function $\theta^* \mapsto \E_G \Psi(\theta^*; W)$ admits a second order Taylor expansion at $\theta$ with a non-singular second derivative matrix $\mathbf{J}(\theta)$.
\item $\Psi(\hat\theta,W) \ge \sup_{\theta^*}\Psi(\theta^*,W) - o_p(n^{-1})$ and $\hat\theta \xrightarrow{p} \theta$.
\end{enumerate}

\setlength{\bibsep}{0pt}
\bibliographystyle{apalike}
\bibliography{./MyLibrary}

\end{document}